\documentclass[aps,prl,twocolumn,superscriptaddress,showpacs]{revtex4}
\usepackage{amsfonts}
\usepackage{amssymb}
\usepackage{amsmath}
\usepackage{graphicx}
\usepackage{dcolumn}
\usepackage{bm}
\usepackage{flafter}

\begin{document}

\title{Anisotropic Dirac fermions in a Bi square net of SrMnBi$_2$}

\author{Joonbum Park}
\affiliation{Department of Physics, Pohang University of Science and Technology, Pohang 790-784, Korea}
\author{G. Lee}
\affiliation{Department of Chemistry, Pohang University of Science and Technology, Pohang 790-784, Korea}
\author{F. Wolff-Fabris}
\affiliation{Dresden High Magnetic Field Laboratory, Helmholtz-Zentrum Dresden-Rossendorf, D-01314 Dresden, Germany}
\author{Y. Y. Koh}
\affiliation{Institute of Physics and Applied Physics, Yonsei University, Seoul 120-749, Korea}
\author{M. J. Eom}
\affiliation{Department of Physics, Pohang University of Science and Technology, Pohang 790-784, Korea}
\author{Y. K. Kim}
\affiliation{Institute of Physics and Applied Physics, Yonsei University, Seoul 120-749, Korea}
\author{M. A. Farhan}
\affiliation{Department of Chemistry, Pohang University of Science and Technology, Pohang 790-784, Korea}
\author{Y. J. Jo}
\affiliation{Department of Physics, Kyungpook National University, Daegu 702-701, Korea}
\author{C. Kim}
\affiliation{Institute of Physics and Applied Physics, Yonsei University, Seoul 120-749, Korea}
\author{J. H. Shim}
\email{jhshim@postech.ac.kr}
\affiliation{Department of Chemistry, Pohang University of Science and Technology, Pohang 790-784, Korea}
\author{J. S. Kim}
\email{js.kim@postech.ac.kr}
\affiliation{Department of Physics, Pohang University of Science and Technology, Pohang 790-784, Korea}
\date{\today}

\begin{abstract}
We report the highly anisotropic Dirac fermions in a Bi square net of SrMnBi$_2$, based on a first principle calculation, angle resolved photoemission spectroscopy, and quantum oscillations for high-quality single crystals. We found that the Dirac dispersion is generally induced in the (SrBi)$^{+}$ layer containing a double-sized Bi square net. In contrast to the commonly observed isotropic Dirac cone, the Dirac cone in SrMnBi$_2$ is highly anisotropic with a large momentum-dependent disparity of Fermi velocities of $\sim 8$. These findings demonstrate that a Bi square net, a common building block of various layered pnictides, provide a new platform that hosts highly anisotropic Dirac fermions.

\end{abstract}


\maketitle
Unlike commonly observed quadratic dispersions, a $linear$ energy dispersion, similar to the spectrum of relativistic Dirac particles, is found in the so-called Dirac materials such as graphene and topological insulators (TIs). Such a linear dispersion is formed when two bands cross each other without hybridization due to the opposite spins or pseudo-spins. Linearly extended bands in Dirac cones, along with the (pseudo)spins, determine the peculiar properties of Dirac materials. For example, back scattering is strongly suppressed leading to a high electron mobility, and anomalous half-integer quantum Hall effects are observed with a nonzero Berry's phase. \cite{review:graphene,review:TI}.

\begin{figure}[b]
\begin{center}
\includegraphics[bb=36 285 524 730, width=7.0cm]{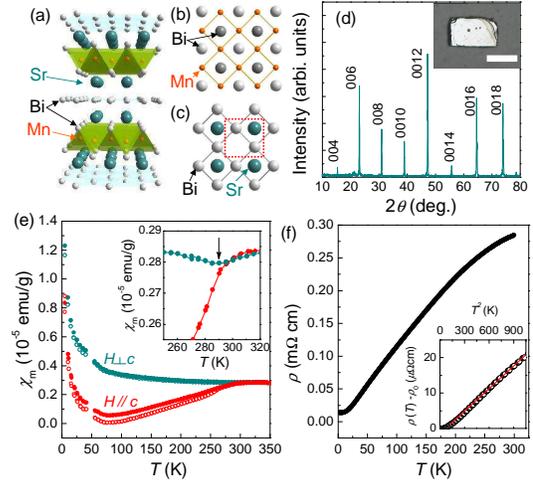}
\caption{\label{fig1} (color online) (a) SrMnBi$_2$ crystal structure. Two building blocks, (b) MnBi layer and (c) Bi square net are shown. Note the double-sized unit cell in the Bi square net. (d) X-ray diffraction pattern from basal planes of a cleaved SrMnBi$_2$ crystal, showing only the (00$l$) reflections. An optical image of a single crystal with a scale of 250 $\mu$m is shown in the inset. (e) Magnetic susceptibilities $\chi(T)$ under $H$ = 1 T applied parallel and perpendicular to $c$-axis in both zero-field-cooled (ZFC, open) and field-cooled (FC, solid) runs. A kink in FC $\chi(T)$ at $T_N$ $\sim$ 290 K is clearly seen in the inset. (f) In-plane resistivity ($\rho$($T$)). It has a quadratic $T$-dependence at low temperatures as shown in the inset.}
\end{center}
\end{figure}

While the Dirac cones in graphene and TIs are in general isotropic ones, different types of Dirac cones have been proposed recently. Highly anisotropic Dirac cone was theoretically predicted in graphene under external periodic potentials\cite{graphene:park:superlattice} or mechanical stress\cite{graphene:son:strain}. Tilted Dirac cones with angle-dependent Fermi velocity have been suggested in a layered organic conductor $\alpha$-(BEDT-TTF)$_2$I$_3$ at high pressures\cite{organic:fukuyama:tiltdirac}. As a hybrid case, a semi-Dirac cone with a quadratic dispersion in one direction and linear in the other may appear in hypothetical graphene with a particular set of hopping parameters between the nearest neighbors\cite{graphene:goerbig:semidirac}. Presence of such a hybrid Dirac cone has been proposed in quantum confined VO$_2$/TiO$_2$ nanostructures.\cite{VO2TiO2:pickett:semiDirac} While the realization of these Dirac fermions may lead to a discovery of novel electronic states $e.g.$ a new-type of quantum Hall state\cite{semidirac:montambaux:QHE}, there has not been a direct experimental observation of above mentioned Dirac Fermions.

In this Letter, we report on a new bulk Dirac material SrMnBi$_2$ that hosts highly anisotropic Dirac fermions in its Bi square net. Based on results from a first principles calculation, quantum oscillations and angle resolved photoelectron spectroscopy (ARPES) on high-quality single crystals, we demonstrate that there is a Dirac dispersion in the electronic structure of the double-sized Bi square net. The Dirac electrons have pseudo-spins stemming from wavefunction amplitudes at two different sublattice, similar to the case of graphene. In contrast to the graphene case, however, the Dirac cone in SrMnBi$_2$ is highly anisotropic showing a significant momentum dependence in the Fermi velocity ($v_F$) with a ratio between the maximum and minimum $v_F$'s of $\sim$ 8. These findings suggest that the Bi square net, a common building blocks of various layered compounds \cite{squarenet:hoffmann:band}, can provide a new platform for highly anisotropic Dirac fermions.

Single crystals of SrMnBi$_2$ were grown by melting stoichiometric mixtures of Sr (99.99$\%$), Mn (99.9$\%$) and Bi (99.999$\%$) chunks in a sealed quartz ampoule. The ampoule was heat-treated at 1150 $^{\rm o}$C, followed by a slow cooling at a rate of 2 $^{\rm o}$C/h. X-ray diffraction and energy dispersive spectroscopy confirm high crystallinity\cite{note_lattice} (see Fig. 1(d)) and right stoichiometry. Magnetotransport properties were measured in a conventional 6-probe configuration in high magnetic fields up to static 33 T at the National High Magnetic Field Laboratory(NHMFL) and up to pulsed 63 T at the Dresden High Magnetic Field Laboratory(HLD). ARPES data were taken with 24 eV photons at the beamline 7U of UVSOR. Samples were cleaved \emph{in situ} in an ultra high vacuum better than $7\times 10^{-11}$ Torr. First principles calculations were done using the full-potential linearized augmented plane wave method\cite{flapw} implemented in WIEN2k package\cite{wien2k}. The generalized gradient approximation (GGA) was utilized for the exchange correlation potential\cite{gga}.

We first look at the basic electronic properties of SrMnBi$_2$. SrMnBi$_2$ consists of a MnBi layer with edge-sharing MnBi$_4$ tetrahedrons and a 2-dimensional (2D) Bi square net stacked with Sr atoms as shown in Fig. 1(a). Due to the low electronegativity of Sr, inserted Sr layers donate charges to the adjacent layers and also electronically separate the MnBi layers and the Bi square net. In the [MnBi]$^-$ layer, Mn$^{2+}$ has a half-filled $3d$ shell in $3d^5$ configuration. The strong Hund coupling of Mn$^{2+}$ leads to a magnetic ground state in Mn$Pn$ ($Pn$ = pnictogen) layers.\cite{note_AFM} Magnetic susceptibility $\chi(T)$ in fact exhibits a kink at $T_N$ $\sim$ 290 K (Fig. 1(e)), indicating an antiferromagnetic (AFM) transition. With the AFM ordering, the charge conduction in MnBi layers is highly suppressed. On the other hand, the covalent nature of Bi 6$p$ bonds in the 2D square net makes the plane conducting, thus governing the transport properties. Resistivity $\rho(T)$ in Fig. 1(f) indeed shows a highly metallic behavior, that is, a quadratic temperature dependence of $\rho(T)$ =  $\rho_0$ + $AT^2$. The parameter $A$, which is inversely proportional to the Fermi temperature, is found to be 19(1) n$\Omega$cmK$^{-2}$, comparable to that of pure Bi\cite{Bi:pratt:res}. This suggests that light carriers are responsible for the metallic conduction.

\begin{figure}
\begin{center}
\includegraphics[bb=45 240 535 740, width=7.5cm]{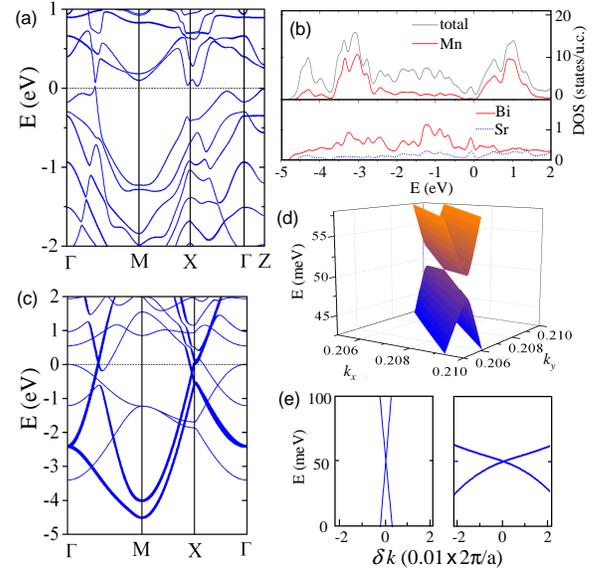}
\caption{\label{fig2} (color online) (a) Band structures of SrMnBi$_2$, (b) total (gray line) and local Mn (red solid) DOS are shown in the upper panel. The lower panel shows local DOSs at Bi (red solid) and Sr (blue dashed) sites in the SrBi layer. (c) Band structure of the isolated (SrBi)$^+$ layer where the line thickness represents the Bi 6$p_{x,y}$ orbital character. (d) Anisotropic energy surfaces around the Dirac point $k_0$ = (0.208, 0.208) for the (SrBi)$^+$ layer. (e) Dispersions near the Dirac point parallel or perpendicular to the $\Gamma$-$M$ symmetry line. SOC in the Bi bands is turned on for the results in (a) and (b), while SOC is not considered for the results in (c), (d) and (e).}
\end{center}
\end{figure}

The electronic structure of SrMnBi$_2$ from a first principles calculation clearly reveals the origin of the light carriers. First of all, from the comparison of the total energies for several magnetic states, we find that a Neel-ordered state in the MnBi layer is the most stable state\cite{note_AFM}. With a Neel-ordering in the MnBi layer, we calculate the band structures of SrMnBi$_2$ with the spin-orbit coupling(SOC) taken into account for the Bi bands (Fig. 2(a)). Due to the large spin polarization of Mn 3$d$ electrons, Mn bands are placed away from the Fermi level ($E_F$) as seen in the density of states (DOS) in Fig. 2(b). The states near $E_F$ are dominated by the Bi states in the square net where the Dirac-like energy dispersion is clearly seen at $k_0$=(0.208, 0.208). A small gap near the Dirac point appears only when the SOC is taken into account, suggesting that this gap is not due to the orbital hybridization but from the SOC. Therefore, the light carriers are, in fact, Dirac fermions residing in the 2D Bi square net.

In order to investigate the nature of the Dirac fermions in the Bi square net, we calculate the band structure of a single (SrBi)$^+$ layer composed of a Bi square net stacked with Sr atoms above and below, and plot it in Fig. 2 (c). For the sake of clarity, we did not include the SOC in this case. We note that Dirac cones exist at several $k$-points but only one at $k_0$=(0.208, 0.208) retains the Dirac dispersion in the band structure of SrMnBi$_2$. Therefore, we focus on the Dirac cone at $k_0$. Near the $E_F$, states have mainly Bi $6p_{x,y}$ character with a small contribution from Sr 4$d$ orbitals due to hybridization. Sr atoms below and above the square net causes unit cell doubling, resulting in two Bi atoms per unit cell. This leads to a folding of the dispersive Bi $6p$ bands. The two Bi 6$p_{x,y}$ bands from two Bi sites cross each other at a single point. There, the pseudo-spin of the Dirac cone in SrMnBi$_2$ can be defined as the amplitude of the wave-function at different Bi sites, similar to graphene.

In spite of the similarity to the graphene, there are differences. First of all, the Dirac cone SrMnBi$_2$ is highly anisotropic as seen in the Dirac dispersion of the (SrBi)$^+$ layer in Fig. 2(d). Along the $\Gamma$-$M$ symmetry line, the Fermi velocity is $v^{\parallel}_F$ = 1.51$\times$10$^6$ m/s which is comparable to that of graphene. In contrast, the dispersion along the cut perpendicular to $\Gamma$-$M$ line is much weaker with $v^{\perp}_F$ of $\sim$ 1.91$\times$10$^5$ m/s, a factor of $\sim$ 8 reduction compared to $v^{\parallel}_F$. The anisotropic nature of the Dirac cone is firmly retained in the full band structure of SrMnBi$_2$. Existence of such anisotropy stems from the fact that the dispersion along the $\Gamma$-$M$ line is determined by overlap between the neighboring Bi atoms in a square net while that along the line perpendicular to the $\Gamma$-$M$ line is due to the hybridization between Sr $d_{xz,yz}$ orbitals and Bi $p_{x,y}$ orbitals. Therefore, the Dirac dispersion in a double-sized Bi square net is always highly anisotropic. Another difference is in the SOC gap. Unlike the negligible SOC gap in graphene\cite{graphene:min:SOC}, there is a sizable SOC gap of $\sim$ 40 meV at the Dirac point in SrMnBi$_2$. This implies that, by placing the Fermi level inside the SOC gap, one can make an $insulating$ Dirac system in which a quantum spin Hall effect is expected to appear\cite{graphene:kane:SHE}. These differences clearly demonstrate that the Dirac cone in the Bi square net of SrMnBi$_2$ is very distinct from that in graphene.

\begin{figure}
\begin{center}
\includegraphics[bb=35 380 476 760, width=8.0cm]{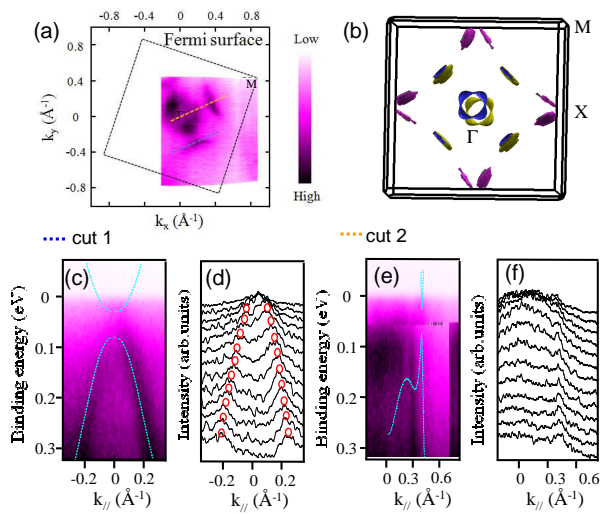}
\caption{\label{fig3} (color online) (a) ARPES intensity map at $E_F$. (b) Calculated FS. (c) Density map and (d) MDCs of the ARPES data along the $\Gamma$-$M$ line (cut 1). (e) and (f) ARPES data along the cut perpendicular to the $\Gamma$-$M$ line (cut 2). The cut directions are shown in panel (a).}
\end{center}
\end{figure}

In order to experimentally investigate the electronic structure, we performed ARPES experiments on SrMnBi$_2$ single crystals. Figure 3(a) shows a FS map which is obtained by integrating the ARPES intensity within 25 meV energy window about the $E_F$. There are two distinct regions with high photoemission intensities: Large circular FS centered at the $\Gamma$ point and a needle-like one between the $\Gamma$ and $M$ points. To compare the data with the calculation, we plot calculated FS's in Fig. 3(b).  There are three small FS's: Needle-shape FS's on the $\Gamma$-$M$ line and near the $X$ point, and flower-shaped hole pockets near the $Z$ point. In comparison with the calculated FS's in Fig. 3(b), the large FS near the $\Gamma$ point should come from the flower-shaped pockets near the $Z$ point as the momentum resolution along the $k_z$ direction is relatively poor due to the finite escape depth of the photoelectrons. The needle-like FS on the $\Gamma$-$M$ line, on which we focus, is consistent with the calculated FS shown in Fig. 3(b).
On the other hand, the calculated FS's near the $X$ point, originated from a band bottom, are not observed. The discrepancy is most likely due to the underestimation of the band gap near X-point in the GGA exchange correlation functional.

Now we focus on the needle-shaped FS on the $\Gamma$-$M$ line where a Dirac dispersion is expected. Figs. 3(c) and 3(e) are ARPES intensity plots along the cuts parallel and perpendicular to the $\Gamma$-$M$ line as indicated in Fig. 3(a). From the raw data and the corresponding momentum distribution curves (MDCs) depicted in Figs. 3(d) and 3(f), energy dispersion with a strong anisotropy is identified. For the cut 1, we mark the dispersion determined from the MDCs in figure 3(d). The experimental band is almost linear at higher binding energies but appear to show some curvature near the $E_F$. Such behavior is consistent with the calculated band with a shift of chemical potential $\sim$ 0.08 eV. in Fig. 3(c) which shows an SOC driven gap. The Fermi velocity $v_F$ is $\sim$ 2 $\times$ 10$^5$ m/s for the cut 1. In a stark contrast, the band for the cut 2 in Fig. 3(e) is extremely dispersive with a much larger $v_F$ ($>$ 1 $\times$ 10$^6$ m/s). In spite of the fact that presence of the SOC does not allow a Dirac point, the overall dispersion shows a Dirac nature with a strong anisotropy.

\begin{figure}
\begin{center}
\includegraphics[bb=25 210 550 775, width=7.5cm]{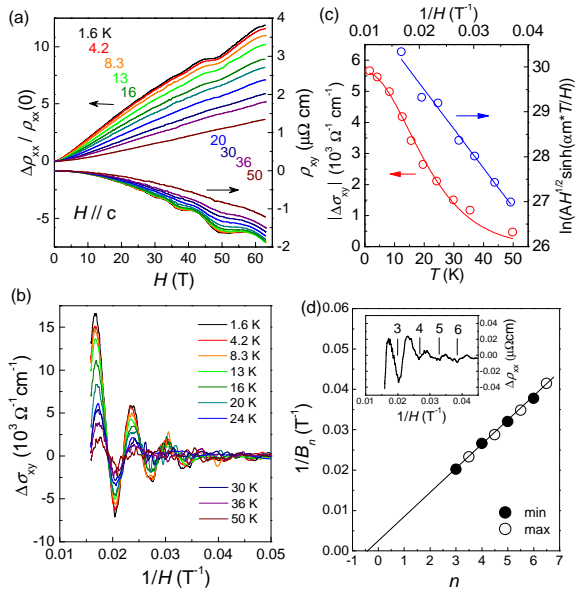}
\caption{\label{fig4} (color online) (a) Magnetic field dependence of $\Delta\rho_{xx}$($H$)/$\rho_0$ and $\rho_{xy}$ of SrMnBi$_2$ (b) Hall SdH oscillations $\Delta \sigma_{xy}$ = $\sigma_{xy}$ - $\langle\sigma_{xy}\rangle$ as a function of 1/$H$ at various temperatures, where $\sigma_{xy}$ = $\rho_{xy}$/($\rho_{xx}^2$+$\rho_{xy}^2$). (c) Cyclotron mass and dingle plots for $\Delta \sigma_{xy}$. (d) SdH fan diagram plotting the measured 1/$B_n$ with the filling factor $n$ which is estimated from the $\Delta \sigma_{xy}$ versus 1/$B_n$ plot shown in the inset.}
\end{center}
\end{figure}

As we have identified an anisotropic linear dispersion at a high energy scale, it will be important to investigate its chiral nature due to pseudo-spins. For that purpose, we have performed magnetotransprot experiments in high magnetic fields. Fig. 4(a) shows the in-plane magnetoresistivity ($\Delta\rho_{xx}$) and Hall resistivity ($\rho_{xy}$) as a function of out-of-plane magnetic field. Clear quantum oscillations in both $\rho_{xx}(H)$ and $\rho_{xy}(H)$ are observed. The background $\rho_{xx}(H)$ exhibits a linear $H$ dependence without any signature of saturation up to 63 T. Such a non-saturating linear magnetoresistance has been often found when the conditions, $n_0$ $\ll$ $(eH/\hbar c)^{3/2}$ and $m^*$ $\ll$ $eH\hbar/cT$, are satisfied\cite{linearMR:abrikosov:thery}. The crossover magnetic field between the regions of $\Delta \rho_{xx}$ $\sim$ $H^2$ and $\Delta \rho_{xx}$ $\sim$ $H$ is only $\sim$ 1 T. This implies that the charge conduction is determined by low-density carriers with an extremely small effective mass, which is consistent with the Shubnikov-de-Hass(SdH) oscillation results in Fig. 4(a). The oscillating component shows a periodic behavior in 1/$B$ with a single frequency of $F$ = 152(5) T. Using the Onsager relation $F$ = ($\Phi_0$/2$\pi^2$)$A_k$, where $\Phi_0$ is the flux quantum and $A_k$ is the cross-sectional area of the Fermi surface normal to the magnetic fields, $A_k$ is found to be 1.45(5) nm$^{-2}$ which corresponds to the $\sim$ 0.7$\%$ of the total area of the Brillouin zone. The $H$ or $T$ dependence of the oscillating amplitude (Fig. 4(b)) gives a cyclotron mass of $m_c$ = 0.29(2) $m_e$ ($m_e$ = free electron mass) and scattering time of $\tau$ = 3.5(5) $\times$ 10$^{-14}$ sec$^{-1}$. The corresponding mobility $\mu$ = $e\tau$/$m_c$ is $\sim$ 250 cm$^2$/Vs. A small FS, small effective mass, and relatively large mobility are consistent with the presence of Dirac fermions.

The key evidence for the existence of Dirac fermions is the non-zero Berry's phase associated with their cyclotron motion. According to a semi-classical magneto-oscillation description, the oscillation part of $\rho_{xx}$ follows $\Delta \rho_{xx}$ $\sim$ $\cos[2\pi(F/B+1/2+\beta]$ where $\beta$ is the Berry's phase (0 $<$ $\beta$ $<$ 1). Experimentally, the Berry's phase can be obtained from the analysis of the SdH fan diagram plotting 1/$B_n$ as a function of the Landau index $n$ which can be determined from the minima of $\rho_{xx}$ as shown in the inset of Fig. 4 (d)\cite{graphene:kim:berry}. In the SdH fan diagram, the intercept of the linear fit yields Berry's phase which is expected to be 1/2 for Dirac fermions in $e.g.$ monolayer graphene. As shown in Fig. 4(d), the linear fit does not go through the origin but rather intercepts the the abscissa at $n$ = -0.40(9). The resulting non-zero Berry's phase reveal that the low-density and high mobile carriers in SrMnBi$_2$ are indeed Dirac fermions.

Based on our findings, we conclude that the Bi square net in SrMnBi$_2$ hosts highly anisotropic Dirac fermions. This implies that SrMnBi$_2$-type compounds containing a Bi square net can be a new platform for anisotropic Dirac fermions with intriguing properties. For example, in $Ae$$T$$Pn_2$ ($Ae$ = alkaline earths, $T$ = transition metals and $Pn$ = pnictogens), the anisotropy of the Dirac cone can be tuned by introducing other alkaline earths, and the SOC gap can also be tuned by replacing Bi with other pnictogens with lower atomic numbers. In addition, since the magnetic ordering of the MnBi layers adjacent to the Bi square net is intimately related to isolation of the Dirac cone from other bands, one could control the coupling between AFM and Dirac fermions by replacing Mn atoms with other transition metals hence modifying the magnetic order. Furthermore, superconductivity can also be incorporated into materials with Dirac fermions as found in a recently-discovered superconductor CeNi$_x$Bi$_2$\cite{CeNiBi2:hosono:SC} and a hypothetical superconductor BaFe$Pn_2$\cite{BaFePn2:kotliar:SC}. Therefore, our findings on Dirac fermions in a layered pnictide, SrMnBi$_2$, containing a square net provide new perspectives not only for controlling the anisotropy of the Dirac cone but also for coupling to magnetic or superconducting orders.

\begin{acknowledgments}
Authors thank Y. W. Son and H. W. Lee for fruitful discussion. This work was supported by LFRIR(2010-00471), BSR (2009-0076700), WCU(R32-2008-000-10180-0) programs through the NRF funded by the MEST. ARPES work was supported by the KICOS (No. K20602000008) and was performed as a joint studie program of IMS in 2010. Experimental work at the HLD was supported by EuroMagNET II (No. 228043). YJJ is supported by the NRF (2010-0006377, 2010-0020057).

\end{acknowledgments}


\begin{thebibliography}{00}
\bibitem{note_phase} The Neel ordering has lower energy than paramagnetic, ferromagnetic, and the stripe-type AFM by 1.64, 0.55, and 0.09 eV/Mn.
\end{thebibliography}
\end{document}